\documentclass[prl,twocolumn,showpacs,preprintnumbers,superscriptaddress,
amsmath,amssymb,tightenlines,epsfig]{revtex4}

\usepackage{bm}
\usepackage{graphicx}

\newcommand{\rv}{{\bf r}}

\newcommand{\ej}{\epsilon_j}
\newcommand{\beq}{\begin{equation}}
\newcommand{\eeq}{\end{equation}}
\newcommand{\bea}{\begin{eqnarray}}
\newcommand{\eea}{\end{eqnarray}}

\newcommand{\<}{\langle}
\renewcommand{\>}{\rangle}

\newcommand{\commentout}[1]{{}}
\newcommand{\half}{\hbox{$1\over2$}}

\renewcommand{\d}{\dag}
\newcommand{\h}{\hat}
\newcommand{\p}{\partial}

\begin{document}

\title{Dissipative quantum dynamics of bosonic atoms in a shallow 1D optical lattice}
\author{J. Ruostekoski}
\affiliation{Department of Physics, Astronomy and Mathematics,
University of Hertfordshire, Hatfield, Herts, AL10 9AB, UK}
\affiliation{Institute for Theoretical Atomic and Molecular Physics,
Harvard-Smithsonian Center for Astrophysics, Cambridge MA 02135}
\author{L. Isella}
\affiliation{Department of Physics, Astronomy and Mathematics,
University of Hertfordshire, Hatfield, Herts, AL10 9AB, UK}
\begin{abstract}
We theoretically study the dipolar motion of bosonic atoms in a very
shallow, strongly confined 1D optical lattice using the parameters
of the recent experiment [Fertig {\it et al.}, Phys.\ Rev.\ Lett.\
{\bf 94}, 220402 (2005)]. We find that, due to momentum uncertainty,
a small, but non-negligible, atom population occupies the unstable
velocity region of the corresponding classical dynamics, resulting
in the observed dissipative atom transport. This population is
generated even in a static vapor, due to quantum fluctuations which
are enhanced by the lattice and the confinement, and is not notably
affected by the motion of atoms or finite temperature.
\end{abstract}

\date{\today}
\maketitle

The rich phenomenology of the dynamics of an atomic Bose-Einstein
condensate (BEC) in a periodic optical lattice potential has
recently attracted considerable theoretical and experimental
interest. In particular, the nonlinear mean-field interaction of the
BEC may give rise to dynamical and energetic instabilities in the
transport properties of the atoms
\cite{BUR01,FER02,WU01,SME02,CAT03,FAL04,CRI04,ZHE04,KOL04,DES04}.
Most experiments to date have been carried out in shallow lattices
where the motion of the BEC can be accurately described by the
classical mean-field theories. In the strongly interacting regime,
with very high lattice potentials, the atoms were shown to undergo a
quantum phase transition to the Mott insulator state
\cite{GRE02a,JAK98}. Recently strongly damped oscillations of
bosonic atoms were experimentally observed by Fertig {\it et al.}
\cite{FER04} at NIST in a tight elongated 1D atom trap, even in the
limit of a very shallow lattice potential. In this Letter we
numerically investigate the damped oscillations of bosonic atoms in
a shallow lattice using the parameters of Ref.~\cite{FER04} and show
that the observed damping rate can be explained by the growing
population in dynamically unstable velocity region, resulting in
phase decoherence. This population appears even without the dipolar
motion and is not notably affected in the studied case by the atom
dynamics. Moreover, for fixed nonlinearity, both, the population at
unstable velocity and the damping rate, have approximately
exponential dependence on the atom number. For larger atom numbers
the quantum fluctuations are suppressed and the finite-temperature
effects can become observable.

In the recent NIST experiment \cite{FER04} an array of independent
1D atom `tubes' were generated by applying a strong transverse 2D
optical lattice potential. The bosonic atoms in each tube were also
trapped along the axial direction by a harmonic magnetic potential
and by a very shallow periodic optical lattice. The dipole
oscillations of atoms along the weak axial lattice were excited by
suddenly displacing the harmonic trap by means of applying a linear
magnetic field gradient. In the absence of the lattice, the dipolar
motion is undamped and unaffected by the interactions. However,
significantly inhibited dynamics was observed even in the presence
of a very shallow lattice with the height of $0.25E_r$, where
$E_r\equiv 2\pi^2\hbar^2 /m\lambda^2$ is the lattice photon recoil
energy with the wavelength $\lambda$. The strong effect of such a
shallow lattice on dissipation is surprising, since the lattice
modulates the atom density only by 6\%. Due to the limited
resolution of the imaging system, it was not possible to measure the
damping of the oscillations in the center-of-mass (cm) position of
the atoms, but the oscillations in the cm velocity were imaged after
a time-of-flight, following the sudden turn-off of the trapping
potentials.

Here we numerically study the dipole oscillations of bosonic atoms
in a 1D lattice within the truncated Wigner approximation (TWA) for
the complete multi-mode field operator, beyond the discrete
tight-binding approximation. The previous theoretical studies of the
quantum dynamics of bosonic atoms in a combined harmonic and lattice
potential indicate that quantum fluctuations can result in strong
dissipation \cite{POL04,ISE04,BAN04,ALT04}. In Ref.~\cite{POL04} the
dynamics was studied within the tight-binding approximation to the
TWA for deep lattices. This approach is much less accurate for very
shallow lattices studied here and in Ref.~\cite{FER04}. Moreover, we
find that the tight-binding approximation to the TWA for shallow
lattices strongly underestimates the generated noise and the effect
of vacuum fluctuations on the dynamics.

In Refs.~\cite{FER04,LAB04} the atoms were confined in a 2D array of
decoupled tight 1D tubes by means adiabatically loading the atoms
into a strong 2D optical lattice. We write the atom density in a 1D
tube at $(x_0,y_0)$:
\beq
\rho(\rv)= {15N_a d^2\over 8\pi^2 l_\perp^2
\bar{R}^3}\big(1-{x_0^2\over R_x^2}-{y_0^2\over R_y^2}-{z^2\over
R_z^2}\big) \, e^{-\bar\rho^2/l_\perp^2}\,, \label{tube}
\eeq
where $\bar\rho^2\equiv(x-x_0)^2+(y-y_0)^2$, $R_i$ are the
Thomas-Fermi radii of the 3D BEC, with
$\bar{R}\equiv(R_xR_yR_z)^{1/3}$, $N_a$ denotes the total atom
number, and $d=\lambda/2$ the lattice spacing. By expanding around
the lattice site minimum we may estimate $l_\perp\simeq
(\hbar/m\omega_\perp)^{1/2}$ by the harmonic trap frequency in the
transverse direction $\omega_\perp =2\sqrt{s} E_r/\hbar$, where $s$
denotes the lattice height in the units of $E_r$. In
Ref.~\cite{FER04}, $s=30$ and $\lambda=810$nm, yielding
$\omega_\perp\simeq2\pi\times 38$kHz. For $N_a=1.4\times10^5$ atoms
\cite{FER04}, $R_x\simeq14\mu m$, $R_y\simeq20\mu m$,
$R_z\simeq10.6\mu m$, and within the classical radius there are $\pi
R_x R_y/d^2\simeq 5400$ atom tubes. From Eq.~(\ref{tube}) we obtain
the number of atoms in each tube $N(x_0,y_0)=\int d^3r \rho=5N_0d^2
(1-x_0^2/R_x^2-y_0^2/R_y^2)^{3/2}/2\pi R_x R_y$. At the central
tube, $N(0,0)\simeq65$ atoms.

When a shallow optical lattice is applied, the wave function
Eq.~(\ref{tube}) is modified along the $z$ axis. In the experiment
the displacement of the harmonic potential was about $\delta\simeq
3\mu$m$\simeq2.2l$, where $l\equiv (\hbar/m\omega)^{1/2}$ and the
trap frequency $\omega\simeq2\pi\times60$Hz. Within $2R_z$, there
are about 52 lattice sites.

We study the NIST experiment of dipolar oscillations within the TWA,
by suddenly displacing the harmonic trap. In the TWA, in a tight
elongated trap $\omega\ll\omega_\perp$, the dynamics of the 1D
classical stochastic field $\psi_W(z,t)$ follows from the nonlinear
equation \cite{Steel}:
\beq
\label{GP} i{\p\over\p t}\psi_W=\big( -{\hbar^2\over 2m}{\p^2\over\p
z^2}+V +g|\psi_W|^2 \big) \psi_W \,,
\eeq
which coincides with the Gross-Pitaevskii equation (GPE). Here the
potential is a combined harmonic trap and a periodic optical lattice
$V(x)=m\omega^2 z^2/2+s E_r \sin^2{(\pi z/d)}$ and
$g=2\hbar\omega_\perp a$, where $a$ denotes the scattering length.
The thermal and quantum fluctuations are included in the initial
state of $\psi_W$ in Eq.~(\ref{GP}) which represents an ensemble of
Wigner distributed wave functions. Our basic formalism of the TWA is
similar to Ref.~\cite{ISE04}. Initially, before displacing the trap,
the gas is assumed to be in thermal equilibrium and we approximate
the field operator $\h\psi(z,t=0)$, within the Bogoliubov theory
\cite{KHE03}:
\beq
\label{field} \h\psi(z)=\psi_0(z)\h\alpha_0+ \sum_{j>0} \big[
u_j(z)\h\alpha_j-v^*_j(z)\h\alpha^\d_j \big]\,.
\eeq
Here $\psi_0$ is the ground state solution of the GPE, obtained by
evolving the GPE in imaginary time, and
$\<\h\alpha_0^\dagger\h\alpha_0\>=N_0$, the number of ground state
atoms. The mode functions $u_j(z)$ and $v_j(z)$ ($j>0$), and the
energies $\epsilon_j$ for excited states are obtained by solving the
corresponding Bogoliubov equations \cite{ISE04} in the harmonic
trap, modulated by the shallow lattice potential. In the TWA the
time evolution of the ensemble of Wigner distributed wave functions
[Eq.~(\ref{GP})] is unraveled into stochastic trajectories, where
the initial state of each realization for $\psi_W$ is generated
according to Eq.~(\ref{field}) with the quantum operators
$(\h\alpha_j,\h\alpha_j^\d)$ replaced by complex
Gaussian-distributed random variables $(\alpha_j,\alpha_j^*)$. These
are obtained by sampling the corresponding Wigner distributions
\cite{ISE04}. In particular, for the quasiparticles the Wigner
function is that of the ideal harmonic oscillators in a thermal bath
\cite{GAR}, with the width $\bar{n}_j+\half$, where $\bar{n}_j\equiv
\<\h\alpha_j^\dagger \h\alpha_j\>= [\exp{(\ej/k_BT)}-1]^{-1}$ and
the $1/2$ width at $T=0$ for each mode represents the quantum noise.

Because the Wigner function returns symmetrically ordered
expectation values, $\<\alpha_j^*\alpha_j\>_W = \bar{n}_j+\half$ and
$\<\alpha_j\>_W=\<\alpha_j^*\>_W=\<\alpha_j^2\>_W=0$, for $j>0$, it
is generally not possible to extract from the TWA simulations the
correlation functions for the full multi-mode field operator
\cite{Steel}. Consequently, we derive the desired normally ordered
expectation values by defining the ground state operators $a_j$ for
each individual lattice site $j$ \cite{ISE04}: $a_j(t)\equiv\int
dz\, \psi^*_{0}(z)\psi_W(z,t)$,
where the integration is over the $j$th lattice site, $\psi_W(z,t)$
is the stochastic field, determined by Eq.~(\ref{GP}), and
$\psi_0(z)$ is the ground state wave function, obtained by evolving
the GPE in imaginary time. This provides us with the basis for
$\tilde\psi(z)=\sum_j \h a_j \phi_j(z)$, where the ground state
functions $\phi_j$ are restricted in the $j$th site, and the
normally ordered expectation values are easily obtained with respect
to $\tilde\psi$, e.g., for the cm position $z_{\rm cm}$, the
position fluctuations $\Delta z=[\<\h z^2\>-\<\h z\>^2]^{1/2}$, and
for the normalized phase coherence between the central well and its
$i$th neighbor $C_i\equiv |\langle \h a^\d_0\h a_i\rangle|
/\sqrt{n_0n_i}$.

\begin{figure}
\includegraphics[width=0.43\columnwidth]{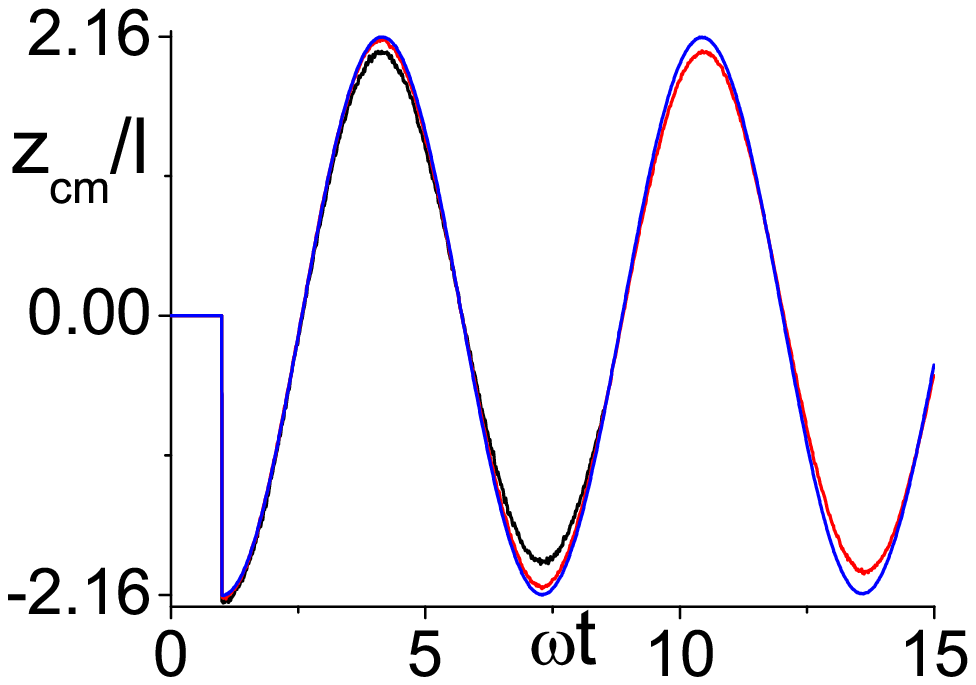}
\includegraphics[width=0.43\columnwidth]{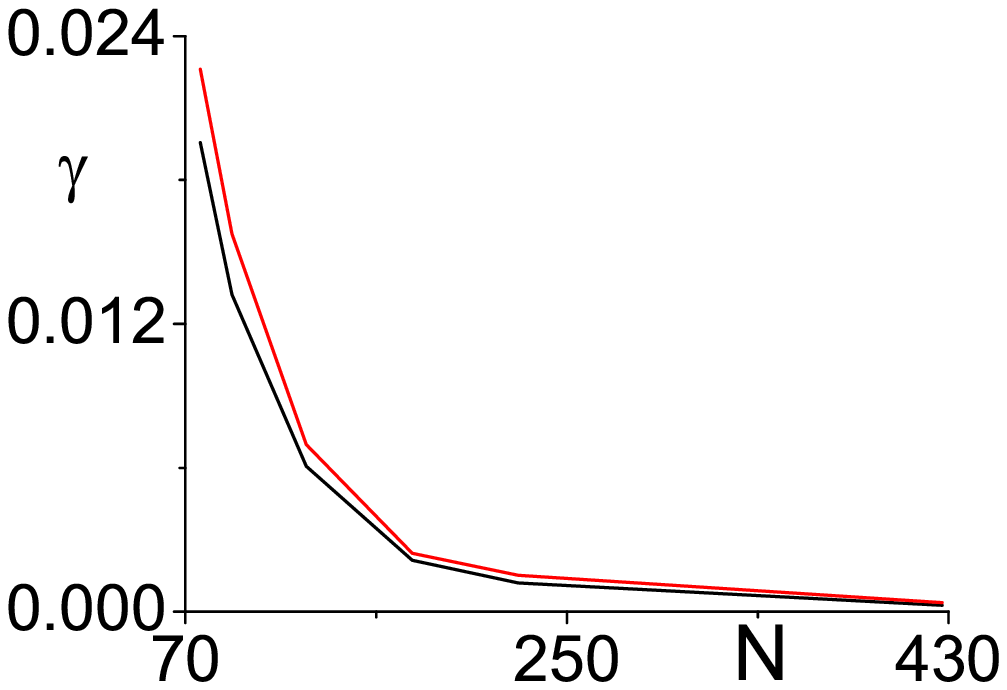}\vspace{-0.5cm}
\includegraphics[width=0.43\columnwidth]{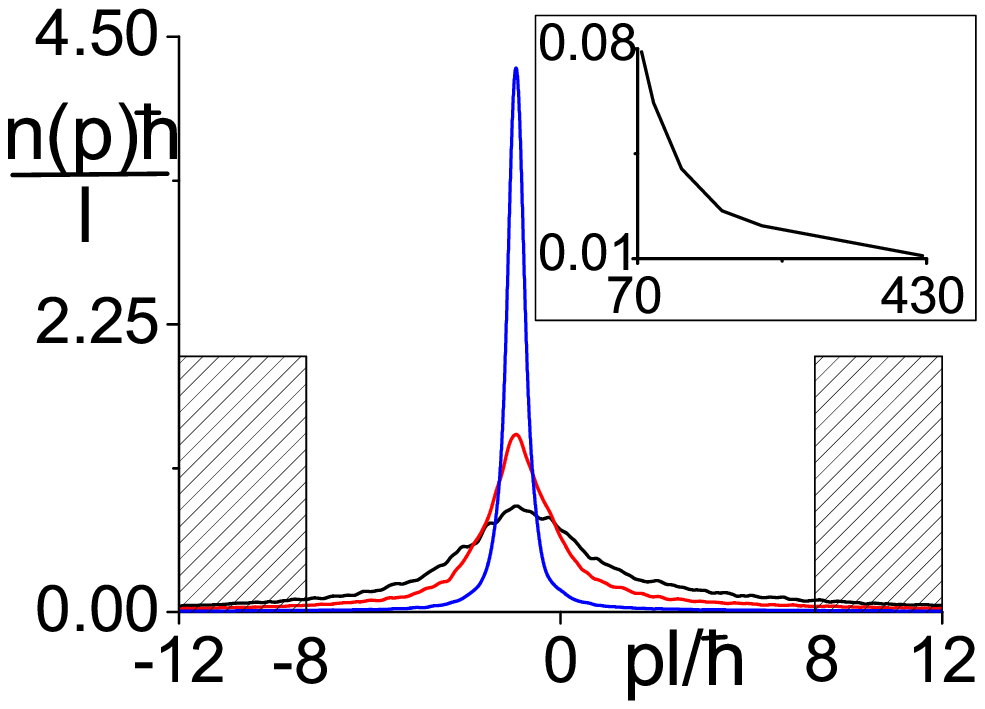}
\includegraphics[width=0.43\columnwidth]{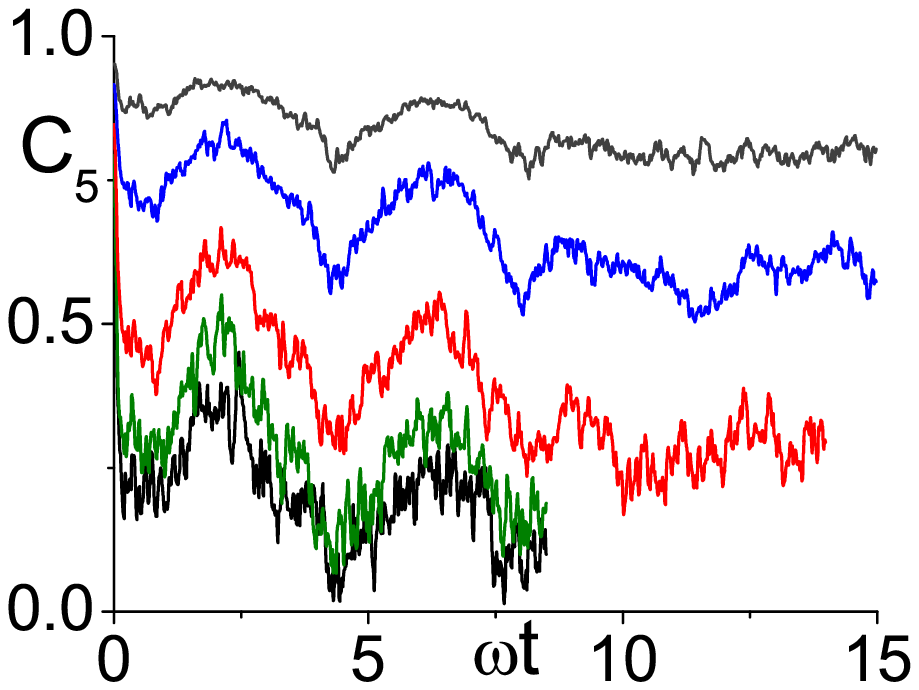}
\vspace{-0.6cm} \caption{The oscillations of the cm position of
atoms from the harmonic trap center at $T=0$ with fixed $Ng$ (top
left; curves from top $N=400,125,75$), the damping rates (top right)
$\gamma(N)$ and $\gamma_v(N)$ (upper line), and the normalized
quasimomentum distributions at $\omega t=8$ (bottom left) with
shaded unstable region; the inset shows the population $n(p)\hbar/
l$ at $v=9\omega l$ where the classical motion becomes unstable. The
trap is instantaneously displaced at $\omega t=1$. The phase
coherence between the initially central well and its fifth neighbor
$C_5$ (bottom right; curves from top $N=1000,400,125,90,75$).
}\label{fig1}
\end{figure}

We numerically study the dipole oscillations using the experimental
parameters \cite{FER04}. In the central atom tube $N\simeq65$,
resulting in the nonlinearity $Ng\simeq320\hbar\omega l$ for
$a\simeq5.313$nm. In Fig.~\ref{fig1} we use the same fixed $Ng$, but
vary $N$, and show the cm position $z_{\rm cm}$ of the atoms for
$s=0.25$. The atoms are initially in thermal equilibrium at $t=0$
and at $\omega t_0=1$ we displace the center of the harmonic trap
from $z=0$ to $z=\delta=2.16l\simeq7.4d$. In the initial state
(\ref{field}), we generate the noise for 300 modes, corresponding to
4-5 lowest energy bands. Even though the atoms mainly remain in the
lowest band, synthesizing the TWA noise for only 70 modes ($\sim$one
band) can underestimate the damping and the phase decoherence
several tens of percents. This emphasizes the importance of the
multi-mode approach to the TWA, beyond the tight-binding
approximation, and is consistent with our previous observations
\cite{ISE04}. In 1D TWA, unlike in 3D \cite{SIN02}, we did not find
any divergence of the calculated results, when the mode number was
increased. For each run, we typically use 600 stochastic
realizations. The integration of the nonlinear evolution in
Eq.~(\ref{GP}) is obtained using the split-step method \cite{JAV04}
on a spatial grid of 4096 points. The numerics becomes much faster
for larger atom numbers, due to smaller quantum fluctuations.

As in Ref.~\cite{FER04}, we model the cm motion as a damped harmonic
oscillator $\ddot{z}_{\rm cm}=-kz/m^*-2\gamma \dot{z}$, so that for
underdamped motion $z_{\rm cm}(t')\equiv -e^{-\gamma t'}(\delta
\cos\Omega t'+c\sin\Omega t')$, with $c\equiv\gamma\delta/\Omega$
and $\Omega=\sqrt{k/m^*-\gamma^2}$, where $t'\equiv t-t_0$ and $m^*$
denotes the effective mass. At $t'=0$ we have $z_{\rm cm}=-\delta$
with $v_{\rm cm}=0$. Even in a very shallow lattice, $s=0.25$, we
find a significant damping for $N=75$ and 90 atoms, with
$\gamma/\omega\simeq0.020$ and 0.013 [Fig.~\ref{fig1}]. The damping
is clearly correlated to the loss of phase coherence. The dependence
of $\gamma$ on $N$ is approximately exponential $\gamma(N)\simeq
0.11\omega \exp{(-N/43)}$, resulting in $\gamma\simeq0.025\omega$ at
the central tube with $N\simeq65$, which is close to the
experimentally observed value of $\gamma\simeq0.03$-$0.04(\pm0.01)
\omega$ \cite{FER04}. We also checked that the damping rate
$\gamma_v$, obtained from $v_{\rm cm}=d z_{\rm cm}/dt$, is close to
$\gamma$. For $s=0.25$ the single-particle effective mass is almost
equal to $m$ and we find $\Omega\simeq\omega$, with very little
effects of fluctuations in $\Omega$.

In the experiment \cite{FER04} the dynamics was measured by imaging
the expanding atom cloud, indicating that $\gamma$ did not represent
the value in the central tube, but the average value over all the
tubes. Although in the experiment $g$ is constant in each tube (as
$N$ varies), we can obtain an upper bound limit for the average
damping rate by extrapolating our fitted $\gamma(N)$ (for fixed
$Ng$) to small $N$, so that for $\gamma/\omega\ll1$,
$\bar\gamma=\sum_{i,j}\gamma(N) N(x_i,x_j)/N_a$. Here $N(x_i,x_j)$
is the atom number in a tube at $(x_i,y_j)$ and the summation is
over all the tubes. By replacing the sum by an integral, we obtain
$\bar\gamma\simeq0.047\omega$.

\begin{figure}
\includegraphics[width=0.43\columnwidth]{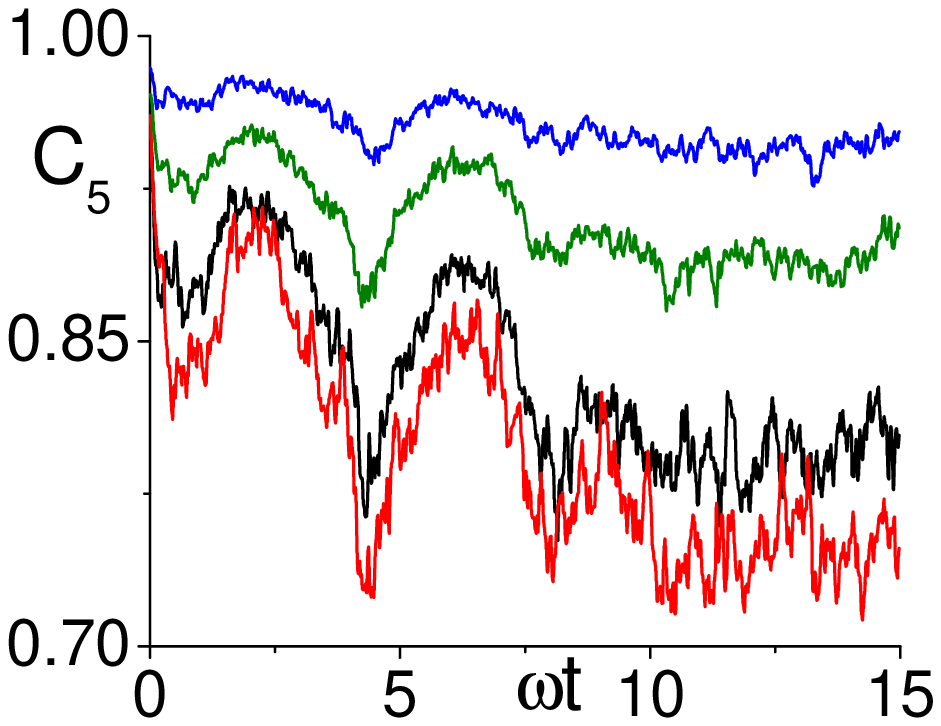}
\includegraphics[width=0.43\columnwidth]{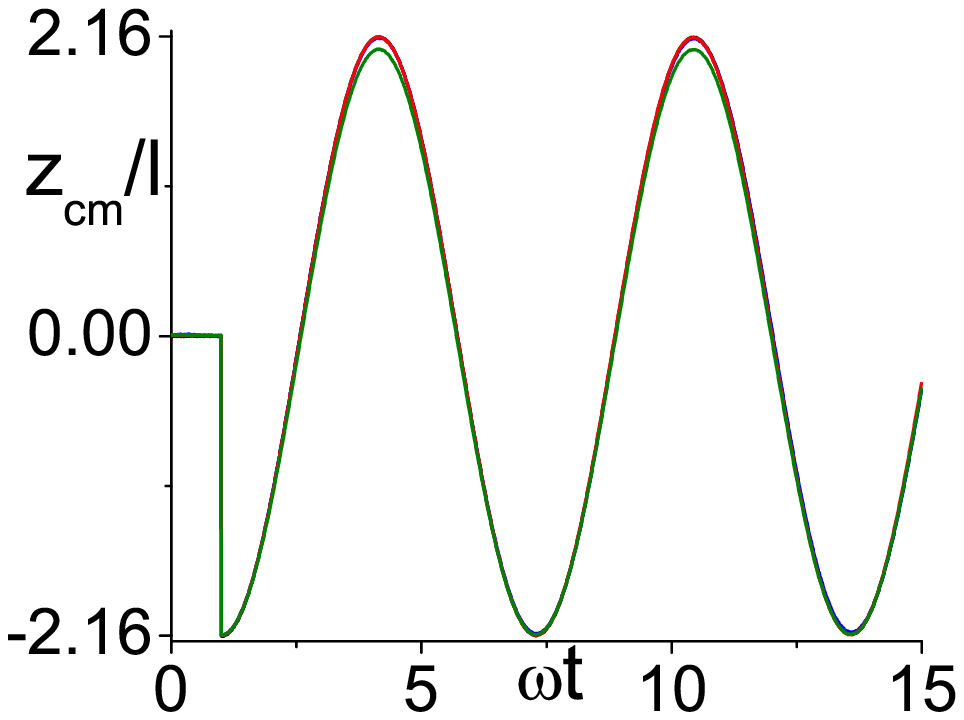}
\vspace{-0.6cm} \caption{Thermal effects on the coherence and the cm
motion. The coherence $C_5$, as in Fig.~\ref{fig1} (left); curves
from top represent ($N_0, k_BT/\hbar\omega)=(1000,0)$, (900,21.3),
(400,0), and (360,10.5). The cm (right) for the same cases. Only the
finite-T, $N_0=900$ case has notable damping
($\gamma\simeq0.002\omega$). } \label{figy}
\end{figure}

We also studied thermal effects on the dipolar motion
[Fig.~\ref{figy}]. In the experiments \cite{FER04} it was not
possible to derive information about temperature and the effect of
thermal atoms on $\gamma$. For small $N$ we found thermal
fluctuations to be negligible compared to quantum fluctuations. Only
for ground state atom numbers clearly exceeding 400, we can
recognize any notable finite-T effects on $\gamma$. This strongly
suggests that the experimentally observed damping \cite{FER04} duly
was due to quantum fluctuations.

The classical nonlinear evolution of a BEC in a periodic lattice is
susceptible to dynamical instabilities arising from both constant
velocity, larger than a critical value
\cite{WU01,SME02,CAT03,FAL04}, and acceleration or force, smaller
than a critical value \cite{CRI04,ZHE04,KOL04}. The onset of the
instability is associated with inhibited transport, fragmentation of
the density profile, and the significant broadening of quasimomentum
distribution. In Ref.~\cite{FER04} the importance of the dynamical
instabilities on dissipation was not clear, since the velocity was
much smaller than the critical value and no momentum broadening was
observed, despite the strongly damped motion.

In the combined harmonic and lattice potential, the separation of
the acceleration and the velocity instabilities is not
straightforward, since the trap induces a local force $m\omega^2 z$
and, e.g., applying a constant force $m\omega^2F$ on atoms is the
same as displacing the trap: $z^2/2-zF=(z-F)^2/2+{\rm const}$.
Moreover, a weakly accelerating lattice does not equal to applying a
constant force, since the acceleration in a shallow lattice
($s=0.25$) does not induce notable cm motion. We numerically
integrated the classical GPE for a BEC (without quantum or thermal
noise) and found the onset of the dynamical instability at
$\delta_c\simeq8$-$9l$, corresponding to a maximum velocity
$v^c_{\rm cm}\simeq8$-$9\omega l$. For smaller displacements, no
instabilities were observed on the time scale of the experiment
\cite{FER04}, e.g., due to the weak force induced by the trap. With
quantum fluctuations, the classically observed sharp onset of the
dynamical instability is smeared out due to the phase uncertainty of
atoms between neighboring lattice sites \cite{POL04}. This is
similar to a BEC in a double-well potential, where the macroscopic
self-trapping of the atom population can decay due to dissipation
\cite{RUO98b}.

In order to investigate the role of the classical dynamical
instabilities in the dissipative quantum dynamics, we calculated the
quasimomentum distribution in the TWA from $\tilde\psi$ and found
atoms occupying the dynamically unstable velocity region of the
corresponding classical system; Fig.~\ref{fig1}. This population is
larger and grows more rapidly, the larger the observed damping:
e.g., at $v=9\omega l$ and $\omega t=8$, $n(p)\hbar/l\simeq0.02$ and
0.08, for $N=225$ and 75, and both $n(p)$ and $\gamma$ show similar
exponential dependence on $N$. For $(N,Ng/\hbar\omega l)=(90,450)$
(here $\gamma\simeq0.026\omega$), the ratio of the population at the
unstable velocity $v\simeq9\omega l$ to that at the peak of the
distribution was 0.07 at $\omega t=3$ and as high as 0.09 at $\omega
t=6$. It is this finite high velocity occupation that results in
dissipation. Alternatively, we may say that the position uncertainty
$\Delta z$ due to vacuum fluctuations is comparable to the critical
displacement $\delta_c$, resulting in dissipation, even though
$\delta<\delta_c$. In the example case $\Delta z\simeq 5$-$6l$. This
is consistent with the observation in Ref.~\cite{BAN04} that the
damping arises from the large depletion, due to the tight transverse
confinement, which leads to the population of high-momentum states.
The tails of the momentum distribution can become very long when the
atoms reach the 1D Tonks-Girardeau regime \cite{OLS98}.

\begin{figure}
\includegraphics[width=0.43\columnwidth]{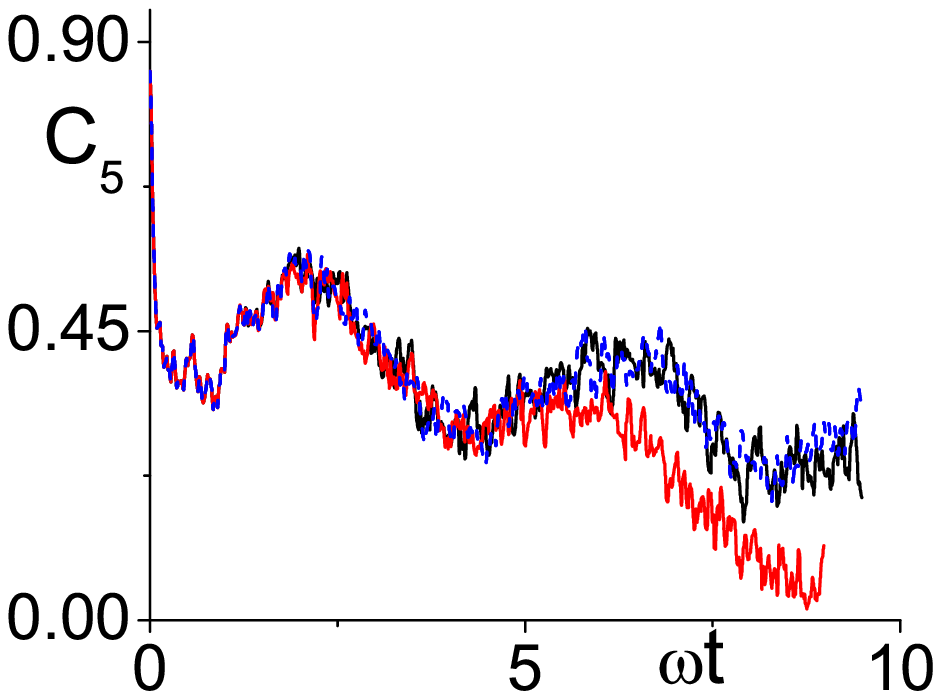}
\includegraphics[width=0.43\columnwidth]{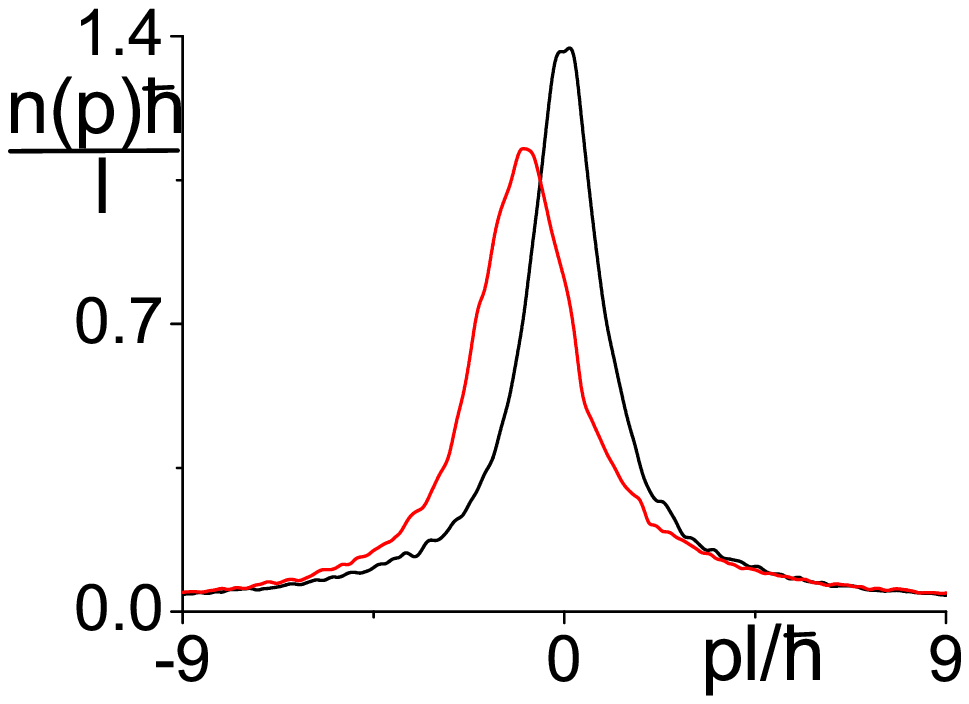}\vspace{-0.6cm}
\caption{The coherence $C_5$ (left) in a displaced trap, in a
lattice moving with a constant velocity $2.16l\omega$, in an
accelerated ($0.27l\omega^2$ and $0.54l\omega^2$) lattice, and in a
static lattice. Only the $0.54l\omega^2$ accelerated lattice
exhibits a lower $C_5$. The quasimomentum distribution (right) in a
displaced trap (off-centered curve) and in a static potential at
$\omega t=6$. } \label{figz}
\end{figure}

Moreover, we also studied the atom response to an accelerated
lattice, to a lattice moving with a constant velocity, and to a
static lattice, without displacing the trap; see Fig.~\ref{figz}. We
used the parameters of Fig.~\ref{fig1}, with $N=125$ and $Ng=405$.
The constant velocity was $v=2.16l\omega$ and the acceleration
$0.27l\omega^2$ over the time period of $\omega t=8$, so that the
final velocity was $2.16l\omega$. In all the cases $z_{\rm cm}$
remained close to zero, without notable cm motion. The decrease in
$C_i$ in all the three systems was almost the same as for the
displaced trap, indicating that the motion of neither the atoms, nor
the lattice, had significant effect on the phase coherence in the
experiments. The occupation of the dynamically unstable region in
the quasimomentum distribution in the static potential was
approximately equal to that in the displaced trap case (e.g., at
$p=\pm 9\hbar/l$, $n\simeq0.05l/\hbar$), resulting in a similar
number of high velocity atoms in the two cases. Moreover, the
variation of this atom number was very little affected by the dipole
oscillations. Consequently, neither phase decoherence nor the
population with unstable velocities, which result in the damping of
the dipole oscillations, are here due to the motion of the atoms,
but the combined effect of the strong transverse confinement and the
lattice. This is contrary to the system studied in Ref.~\cite{POL04}
where the atom motion itself was argued to generate the decoherence.

As expected, both the increase of the lattice height and the initial
displacement enhance dissipation. We studied the dynamics for $s=1$,
$\delta=2.16l$, and obtained for $(N,Ng/\hbar\omega
l)=(330,350),(430,340),(830,330)$, $\gamma/\omega\simeq
0.019,0.012,0.003$, at $T=0$. Also with $s=1$ the damping is notable
for $\delta$ much smaller than the classical critical displacement
$\delta_c\simeq 6$-$7l$. On the other hand, for $N=175$ and
$Ng\simeq380\hbar\omega l$, the increase of $\delta/l$ from $2.16$
to $4.32$ increased $\gamma/\omega$ from $0.004$ to $0.009$.

In summary, we studied dissipative atom transport in a shallow 1D
lattice. At a fixed nonlinearity, smaller $N$ results in larger
phase and momentum uncertainty and, consequently, larger population
in the dynamically unstable velocity region. Surprisingly, this
population is generated due to the confinement and the lattice, even
with neither the atoms nor the lattice moving. The qualitative
agreement between the TWA and the experiment \cite{FER04}, despite
the approximations involved in generating the initial state at small
$N$, also represents a considerable success of the TWA which is
traditionally considered more suitable in the limit of large field
amplitude \cite{Steel,SIN02}. For instance, the time-dependent
Hartree-Fock-Bogoliubov theory predicts much too small a damping
\cite{BAN04}. This suggests the TWA could provide a powerful
technique to study vacuum fluctuations also in several other related
condensed matter and atomic systems, such as corrugated
superconducting nanowires \cite{BEZ00}.

\acknowledgments{We acknowledge financial support from the EPSRC and
NSF through a grant for the Institute for Theoretical Atomic,
Molecular and Optical Physics at Harvard University and Smithsonian
Astrophysical Observatory. After our work was completed, the
dynamics in a deep lattice was analyzed in the fermionized limit
\cite{REY05}, in the Mott state. In this limit the TWA with a
bosonic field amplitude is no longer valid.}

\end{document}